\pdfoutput=1
\documentclass[aps,pra,twocolumn,superscriptaddress,floatfix]{revtex4-2}
\usepackage[a4paper,textwidth=183 mm,textheight = 247mm]{geometry}
\usepackage[dvipsnames]{xcolor}
\usepackage{graphicx}
\usepackage[tbtags]{amsmath}
\usepackage{amsthm}
\usepackage{empheq}
\usepackage{amssymb}
\usepackage{mathrsfs}
\usepackage{float}
\usepackage{array}
\setcitestyle{super,open={},close={},citesep={,}}

\usepackage{caption}
\DeclareCaptionLabelSeparator{line}{\,\textbf{\textbar}\,}
\usepackage{color}
\usepackage{dcolumn}
\usepackage{bm}
\usepackage{here}
\usepackage{braket}
\allowdisplaybreaks

\newcommand{\mean}[1]{\langle#1\rangle}
\newcommand{\abs}[1]{\left\vert#1\right\vert}

\begin{document}

\title{Propagating Gottesman-Kitaev-Preskill states encoded in an optical oscillator}
\author{Shunya Konno} 
\affiliation{Department of Applied Physics, School of Engineering, The University of Tokyo, 7-3-1 Hongo, Bunkyo-ku, Tokyo 113-8656, Japan}

\author{Warit Asavanant} 
\email{warit@alice.t.u-tokyo.ac.jp} 
\affiliation{Department of Applied Physics, School of Engineering, The University of Tokyo, 7-3-1 Hongo, Bunkyo-ku, Tokyo 113-8656, Japan}
\affiliation{Optical Quantum Computing Research Team, RIKEN Center for Quantum Computing,\\ 2-1 Hirosawa, Wako, Saitama 351-0198, Japan}

\author{Fumiya Hanamura} 
\affiliation{Department of Applied Physics, School of Engineering, The University of Tokyo, 7-3-1 Hongo, Bunkyo-ku, Tokyo 113-8656, Japan}

\author{Hironari Nagayoshi} 
\affiliation{Department of Applied Physics, School of Engineering, The University of Tokyo, 7-3-1 Hongo, Bunkyo-ku, Tokyo 113-8656, Japan}

\author{Kosuke Fukui} 
\affiliation{Department of Applied Physics, School of Engineering, The University of Tokyo, 7-3-1 Hongo, Bunkyo-ku, Tokyo 113-8656, Japan}

\author{Atsushi Sakaguchi} 
\affiliation{Optical Quantum Computing Research Team, RIKEN Center for Quantum Computing,\\ 2-1 Hirosawa, Wako, Saitama 351-0198, Japan}

\author{Ryuhoh Ide} 
\affiliation{Department of Applied Physics, School of Engineering, The University of Tokyo, 7-3-1 Hongo, Bunkyo-ku, Tokyo 113-8656, Japan}

\author{Fumihiro China}
\affiliation{Advanced ICR Research Institute, National Institute of Information and Communications Technology, 588-2 Iwaoka, Nishi, Kobe 651-2492, Japan}

\author{Masahiro Yabuno}
\affiliation{Advanced ICR Research Institute, National Institute of Information and Communications Technology, 588-2 Iwaoka, Nishi, Kobe 651-2492, Japan}

\author{Shigehito Miki}
\affiliation{Advanced ICR Research Institute, National Institute of Information and Communications Technology, 588-2 Iwaoka, Nishi, Kobe 651-2492, Japan}
\affiliation{Graduate School of Engineering, Kobe University, 1-1 Rokko-dai, Nada, Kobe 657-0013, Japan}

\author{Hirotaka Terai}
\affiliation{Advanced ICR Research Institute, National Institute of Information and Communications Technology, 588-2 Iwaoka, Nishi, Kobe 651-2492, Japan}

\author{Kan Takase}
\affiliation{Department of Applied Physics, School of Engineering, The University of Tokyo, 7-3-1 Hongo, Bunkyo-ku, Tokyo 113-8656, Japan}
\affiliation{Optical Quantum Computing Research Team, RIKEN Center for Quantum Computing,\\ 2-1 Hirosawa, Wako, Saitama 351-0198, Japan}

\author{Mamoru Endo}
\affiliation{Department of Applied Physics, School of Engineering, The University of Tokyo, 7-3-1 Hongo, Bunkyo-ku, Tokyo 113-8656, Japan}
\affiliation{Optical Quantum Computing Research Team, RIKEN Center for Quantum Computing,\\ 2-1 Hirosawa, Wako, Saitama 351-0198, Japan}

\author{Petr Marek}
\affiliation{Department of Optics, Palacky University, 17. listopadu 1192/12, 77146 Olomouc, Czech Republic}

\author{Radim Filip}
\affiliation{Department of Optics, Palacky University, 17. listopadu 1192/12, 77146 Olomouc, Czech Republic}

\author{Peter van Loock}
\affiliation{Institute of Physics, Johannes-Gutenberg University of Mainz, Staudingerweg 7, 55128 Mainz, Germany}

\author{Akira Furusawa} \email{akiraf@ap.t.u-tokyo.ac.jp}\affiliation{Department of Applied Physics, School of Engineering, The University of Tokyo, 7-3-1 Hongo, Bunkyo-ku, Tokyo 113-8656, Japan}
\affiliation{Optical Quantum Computing Research Team, RIKEN Center for Quantum Computing,\\ 2-1 Hirosawa, Wako, Saitama 351-0198, Japan}

\maketitle

\textbf{A quantum computer with low-error, high-speed quantum operations and capability for interconnections is required for useful quantum computations. A logical qubit called Gottesman-Kitaev-Preskill (GKP) qubit \cite{PhysRevA.64.012310} in a single Bosonic harmonic oscillator is efficient for mitigating errors in a quantum computer. The particularly intriguing prospect of GKP qubits is that entangling gates as well as syndrome measurements for quantum error correction only require efficient, noise-robust linear operations. To date, however, GKP qubits have been only demonstrated at mechanical and microwave frequency in a highly nonlinear physical system \cite{Fluhmann2019,Campagne-Ibarcq2020}. The physical platform that naturally provides the scalable linear toolbox is optics, including near-ideal loss-free beam splitters and near-unit efficiency homodyne detectors that allow to obtain the complete analog syndrome for optimized quantum error correction \cite{PhysRevLett.119.180507}. Additional optical linear amplifiers \cite{PhysRevLett.113.013601,PhysRevA.98.052311} and specifically designed GKP qubit states are then all that is needed for universal quantum computing \cite{PhysRevResearch.3.043026,PhysRevLett.123.200502,PhysRevResearch.2.023270}.  In this work, we realize a GKP state in propagating light at the telecommunication wavelength and demonstrate homodyne meausurements on the GKP states for the first time without any loss corrections. Our GKP states do not only show nonclassicality and non-Gaussianity at room temperature and atmospheric pressure, but unlike the existing schemes with stationary qubits, they are realizable in a propagating wave system. This property permits large-scale quantum computation \cite{Asavanant373,Larsen369,doi:10.1063/1.4962732,Yokoyama2013,doi:10.1126/sciadv.abj6624} and interconnections \cite{PhysRevResearch.3.033118,Rozpedek2021}, with strong compatibility to optical fibers and 5G telecommunication technology \cite{10.1063/5.0137641}.}

A quantum computer operates based on quantum mechanics which allows it to outperform classical computers in certain tasks. In the recent years, there were a few first demonstrations of quantum supremacy, the capability to go beyond classical computation, on various platforms \cite{Arute2019,PhysRevLett.123.250503,doi:10.1126/science.abe8770}. It is beyond any doubt that within a few years, noisy intermediate-scale quantum computers will be available. To go beyond that, however, fault-tolerant quantum processors are required. In order to mitigate the errors in the quantum computation, quantum error correction (QEC) is currently the most prominent strategy. QEC allows detections and corrections of errors without disturbing the quantum information by having additional redundancy in the system. QEC using Bosonic systems has attracted attention due to their large Hilbert spaces capable of redundantly encoding qubits, and various Bosonic codes such as binomial codes \cite{PhysRevX.6.031006}, cat codes \cite{PhysRevA.59.2631,PhysRevLett.111.120501}, and Gottesman-Kitaev-Preskill (GKP) codes \cite{PhysRevA.64.012310} are being explored.

In terms of wave functions, GKP states have a periodic sharp-peaked grid structure in both of their quadratures, allowing for detections and corrections of small loss and displacement errors which are dominant in the optical systems. Advantageously, the basic qubit-level operations on the GKP codes (also called Clifford operations), including entangling gates and error syndrome measurements, can be done using only linear operations. Moreover, universal quantum computation can be achieved by additional light squeezers \cite{PhysRevLett.113.013601,PhysRevA.98.052311}, which have been already demonstrated in optical systems, and specifically designed GKP qubit states \cite{PhysRevResearch.3.043026,PhysRevLett.123.200502,PhysRevResearch.2.023270}. The key technologies for fault-tolerant quantum computation using GKP qubits are based on quantum teleportation \cite{PhysRevLett.112.120504}. Quantum teleportation providses the capability for linear operations with Bell measurement and feedforward displacement corresponding to syndrome measurement and recovery operation \cite{,PhysRevA.104.062427,PhysRevA.71.042322,PhysRevA.107.032412}. Quantum teleportation technology is pioneered and highly developed in the optical system \cite{Furusawa706,Lee330,PhysRevApplied.16.034005,Larsen2021}, making it a promising platform for GKP states. The experimental demonstrations of the GKP qubits up to date, however, are in stationary systems that can be easily coupled to qubits \cite{Fluhmann2019,Campagne-Ibarcq2020} which provides strong nonlinearity. On the other hand, linear operations are not naturally available and have to be constructed by engineering the nonlinearity of the system \cite{Gao2019,PhysRevX.13.021004}, limiting the scalability to a large-scale multimode interaction. This is in contrast to the optical system where linear operations can be simply done with commercially available beamsplitters and multimode linear operations can be easily implemented \cite{PhysRevApplied.16.034005,Larsen2021}.

\begin{figure*}[htp]
\includegraphics[width=\textwidth]{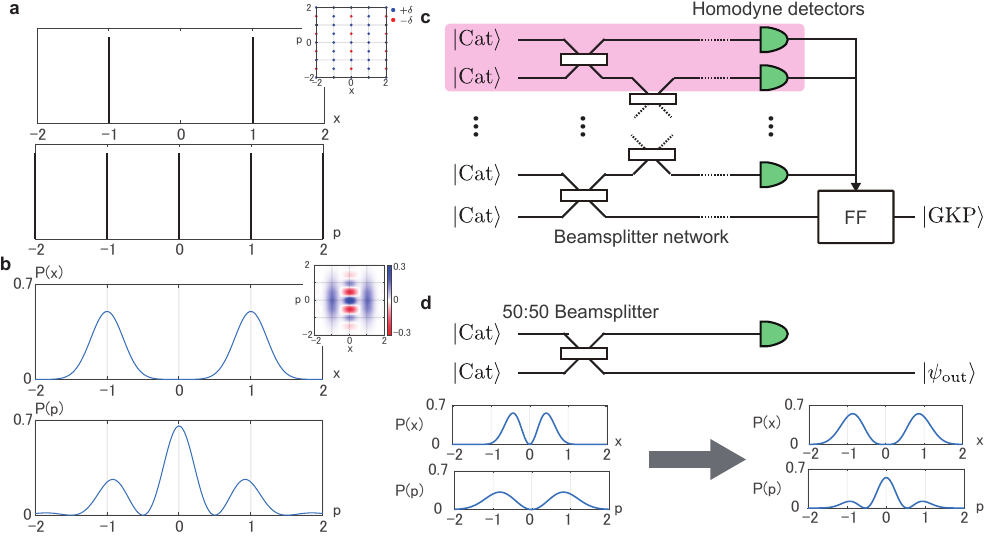}%
\caption{GKP state and its generation method in the optical system. \textbf{a,b,} Quadrature distributions of ideal (infinite energy) and approximated GKP $\ket{1}$ states with 5 dB of squeezing, respectively (the units are defined with ${\protect\hbar}=1$ and the quadratures are normalized to $\sqrt{\pi}$). The insets are the Wigner functions. \textbf{c,} General generation method in the optical system based on the intereference of cat states using a beamsplitter network and implementing homodyne measurements and feedforward operations (FF) incorporating squeezing and displacement operations. The highlighted part is the single step of the generation method. \textbf{d,} Single step of the generation method. Note that FF with Gaussian operations are omitted in this figure and are replaced by conditioning in this experiment. One can see the formation of peaks at the stabilizer values.\label{fig:schematic}}
\end{figure*}

In addition to the linearity of the system, there are a few key properties required for the GKP states to be useful in actual quantum computation. The physical platform for using GKP states should allow large-scale and fast operations, as slow implementation of syndrome measurements and corrections increases the error rates and thus entails an additional overhead. Moreover, for applications such as quantum communication or interconnecting quantum computers using quantum internet, GKP states encoded in a propagating wave can be the core elements. From such aspects, a propagating electromagnetic wave system in the optical regime is one of the promising candidates. With multiplexing of degree of freedom such as time \cite{PhysRevA.83.062314,Asavanant373,Larsen369,doi:10.1063/1.4962732,Yokoyama2013,doi:10.1126/sciadv.abj6624} or frequency \cite{PhysRevLett.112.120505,PhysRevA.94.032327,Pfister_2020} in the propagating wave, a large-scale quantum computation platform have already been demonstrated. Also, terahertz-bandwidth light source \cite{10.1063/5.0063118}, 43-GHz optical homodyne measurement \cite{10.1063/5.0137641}, and high-speed nonlinear feedforward \cite{2022arXiv221017120S}---key components to high-speed optical quantum computation and error correction---have been demonstrated, meaning that we can expect near-term optical quantum computation with a clock frequency of at least a few gigahertz, surpassing other physical systems by a several orders of magnitude. Despite these appealing features, however, the actual optical generation of GKP qubit in propagating optical system has remained elusive as propagating electromagnetic systems lack viable strong nonlinearity, and even if we try to obtain nonlinearity via a system such as cavity QED \cite{Magro2023}, a complex arrangement would be required to realize a complex quantum state. By realizing GKP states in the optical system, we can overcome the limitations of multimode linear operations and scalability of the GKP approach in the nonlinear systems \cite{Fluhmann2019,Campagne-Ibarcq2020}. Therefore, realization of GKP states in the propagating wave is a key to practical quantum computation and is the final main ingredient of the fault-tolerant universal quantum computer using optical systems.

In this work, we achieve the first generation of the GKP state in a propagating wave system. Our generation method is based on the two-mode interference between cat states and a single-mode projection via an optical homodyne measurement \cite{Vasconcelos:10,PhysRevA.97.022341}. The required nonlinearity in the GKP state generation is introduced off-line via photon number measurements used in the generation of the cat states. The generated state is characterized via homodyne measurements and is reconstructed with quantum tomography. No corrections for experimental imperfection are used in both homodyne measurements and quantum tomography. Homodyne measurement is a linear measurement which is required in the syndrome measurements and operations of GKP states. Homodyne measurement on the GKP states, however, has not been implemented in the previous experimental demonstrations \cite{Fluhmann2019,Campagne-Ibarcq2020}. Although the cat state generation in this work is probabilistic, by combination with the cutting-edge photon number resolving detector \cite{Endo:23} and methods for high-rate cat state generation \cite{PhysRevA.103.013710}, the whole process can have high success rate. Also, for a more complex multistep generation, feedforward displacement based on the results of the homodyne measurements can be done to remove the necessity of the homodyne conditioning \cite{,PhysRevA.97.022341}, making the process after the cat state generation semi-deterministic. In addition, the GKP states generated in this work are at the telecommunication wavelength of 1545 nm which makes the generated states highly compatible with optical fibers and 5G technology in the telecommunication. Therefore, our generation of GKP state in a propagating wave will be a basis for fault-tolerant quantum computation \cite{PhysRevLett.112.120504} as well as quantum communication \cite{PhysRevResearch.3.033118,Rozpedek2021}.

%

The GKP state can be defined in several ways. When we consider the basic observables of light, quadrature operator $\hat{x}$ and $\hat{p}$ ($[\hat{x},\hat{p}]=i$), the Pauli operators on the phase-space grid are $\hat{\bar{X}}=\exp(-i\sqrt{\pi}\hat{p})$ and $\hat{\bar{Z}}=\exp(i\sqrt{\pi}\hat{x})$ which corresponds to displacement in $x$ and $p$ by $\sqrt{\pi}$ respectively. Then, the GKP states are defined in the logical space that is stabilized, i.e. invariant, under the operators $\hat{S}_x=\hat{\bar{X}}^2=\exp(-i2\sqrt{\pi}\hat{p})$ and $\hat{S}_p=\hat{\bar{Z}}^2=\exp(i2\sqrt{\pi}\hat{x})$ which are displacements by $2\sqrt{\pi}$ in the $x$ and $p$ direction, respectively \cite{PhysRevA.64.012310}. Alternatively, for any ideal GKP states in the logical space, we have $\langle\hat{S}_x\rangle=\langle\hat{S}_p\rangle=1$. This means that ideal GKP states should have periodicity of $2\sqrt{\pi}$ in both $x$ and $p$. Defining these two stabilizers does not uniquely determine the states, but they determine a logical space, in which the information is encoded, and we have to consider another stabilizer to define our state. We can define our basis for the logical space, i.e., the eigenstates for $\hat{\bar{Z}}$, as $\ket{0_L}\propto\sum_{k}\ket{x=2k\sqrt{\pi}}$ and $\ket{1_L}\propto\sum_{k}\ket{x=(2k+1)\sqrt{\pi}}$. These two states have $\hat{\bar{Z}}$ and $-\hat{\bar{Z}}$ as their stabilizers, respectively, making one of the two GKP code stabilizers $\hat{S}_p$ redundant. Note that $\pm\hat{\bar{Z}}$ commutes with both $\hat{S}_x$ and $\hat{S}_p$. As these states are unnormalizable, they are usually approximated by replacing the position eigenstates with squeezed states and the whole states are enveloped with a Gaussian envelope for the symmetry in both $x$ and $p$ quadrature \cite{PhysRevA.102.032408}. Figure \ref{fig:schematic}(a,b) shows the grid quadrature distribution of an ideal and approximated GKP state.

\begin{figure*}[htb]
\includegraphics[width=\textwidth]{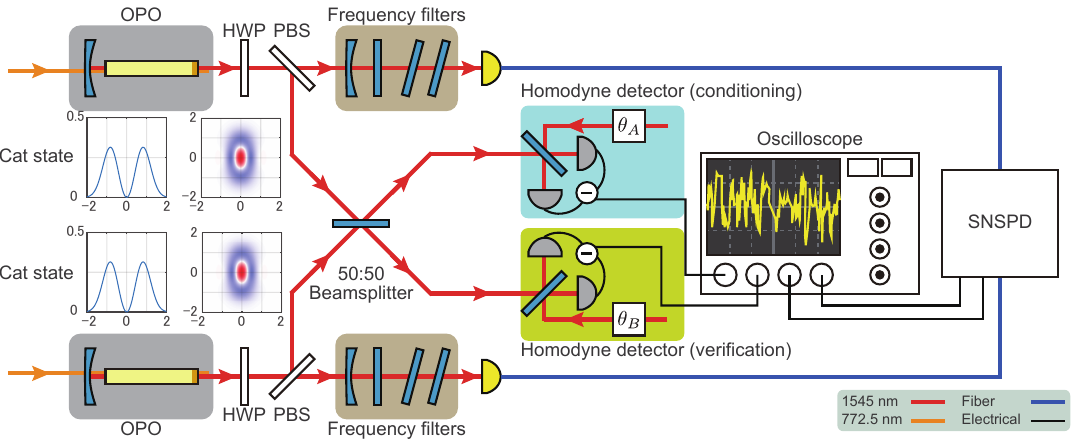}%
\caption{Experimental setup. OPO, optical parametric oscillator; HWP, half-wave plate; PBS, polarization beamsplitter; SNSPD, superconducting nanostrip single photon detector. The insets show the ideal states from the photon subtraction with the quadrature distribution in $p$. The phases of the two homodyne measurements are given by $\theta_A$ and $\theta_B$, where $\theta_A$ is set to $0^\circ$ for measuring and conditioning of $x$ quadrature and $\theta_B$ is set to $0^\circ$, $30^\circ$, $60^\circ$, $90^\circ$, $120^\circ$, $150^\circ$ for collecting quadrature values for quantum tomography.\label{fig:experimental_setup}}
\end{figure*}
We generate the peak structure of the GKP states by using the interference of kitten states \cite{Ourjoumtsev83,PhysRevLett.97.083604,Wakui:07,Asavanant:17} and homodyne measurements \cite{Vasconcelos:10,PhysRevA.97.022341}. Figure \ref{fig:schematic}c shows a diagram of the general generation method. For the intuition, let us consider a single interference step (shown in Fig.\ \ref{fig:schematic}d). If we interfere two cat states $\ket{\psi_\textrm{cat}}\propto\ket{i\alpha}-\ket{-i\alpha}$, where $\ket{i\alpha}$ and $\ket{-i\alpha}$ are coherent states with $\alpha$ being a real number, the output two-mode state is $(\ket{i\sqrt{2}\alpha}+\ket{-i\sqrt{2}\alpha})\ket{0}-\ket{0}(\ket{-i\sqrt{2}\alpha}+\ket{i\sqrt{2}\alpha})$. If we measure the first mode and condition it at $x=0$, we will approach $\ket{\psi_\textrm{out}}\propto\ket{-i\sqrt{2}\alpha}-2\ket{0}+\ket{i\sqrt{2}\alpha}$. The output state now approaches the central part of GKP state in the phase space (up to additional squeezing), and the process can be repeated so that a larger part of the GKP state is synthesized. The coefficients of the peaks in this method can be easily shown to be binomial coefficients which approach Gaussian distribution for a large number of peaks \cite{Vasconcelos:10}. To make the width of the peaks small and the distance between peaks correct as a GKP state, we either implement the squeezing operation at the end, or start with squeezed cat states instead of a cat state \cite{Vasconcelos:10,PhysRevA.97.022341}. Such Gaussian operations only shape the GKP state but do not increase the non-Gaussian grid aspects. Although we use ordinary cat states for our experimental demonstration, squeezing operation on cat state \cite{PhysRevLett.113.013601} and high-rate generation of squeezed cat states \cite{PhysRevA.103.013710} are both being explored in optics. 
%

Figure \ref{fig:experimental_setup} shows the experimental system. The master laser of the system is a continuous-wave laser with a wavelength of 1545.32 nm with a second harmonic generator for generation of 772.66 nm light. The kitten states for the interference are generated by using photon subtraction on squeezed light \cite{PhysRevA.55.3184,Takase:22,Asavanant:17,PhysRevA.82.031802}. The squeezed light sources are optical parametric oscillators (OPOs) whose design is based on Ref.\ \cite{Takanashi:19}. The detection of the photon in the photon subtraction is done by the superconducting nanostrip single photon detectors (SNSPDs) \cite{Miki:17}. The coincidence detection at both SNSPDs herald the success in the interference. We measure one of the modes in the $x$-basis using a homodyne detector, so that the GKP state is generated in the other mode. The signal of the SNSPDs acts as the measurement trigger for the oscilloscope and the electrical signals of homodyne detectors are measured in real time. To perform quantum tomography, we measure the quadrature of the other mode at various phases and use the collected data to reconstruct the state via maximum likelihood method \cite{Lvovsky_2004}. No correction for optical losses and any experimental imperfections are done on the measurement results or in the tomography process.
%

\begin{figure*}[htb]
\includegraphics[width=\textwidth]{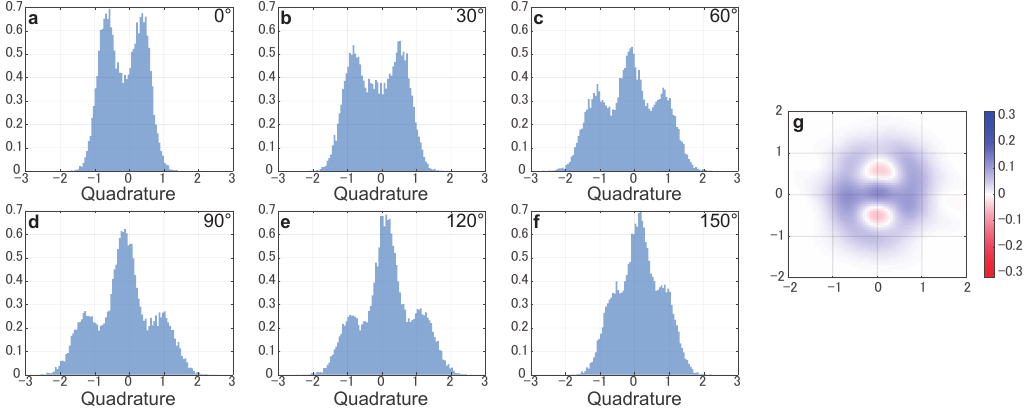}%
\caption{Histograms of the experimentally obtained quadrature values of the generated states and their reconstructed Wigner function. The distributions are normalized as a probability density function. No corrections are done on the obtained distributions. \textbf{a--f,} Quadrature distributions of the generated states. The homodyne phases are given in each subfigure. \textbf{g,} The reconstructed Wigner function. The appropriate adjustments of the fixed linear operations on the output state are implemented (see supplementary).\label{fig:quadrature}}
\end{figure*}
Figure \ref{fig:quadrature} shows the measured quadrature distribution in each phase of the generated state and the reconstructed Wigner functions. Initially, the kitten state has only two peaks on the both quadratures. By implementing the protocol, the structure changes to the central part of the GKP state and we can see qualitative resemblance between the quadrature distribution of the experimentally obtained data to the ideal case in Fig.\ \ref{fig:schematic}. The overlaps between the peaks are the results of the experimental imperfection such as optical losses. From the quadrature distributions, we reconstruct the generated states. We observe two regions with negative values. These synthesized negative regions are the evidence for the merging of the nonclassical aspects of the kitten states to the GKP grid structure.

For GKP states to be useful, they have to be able to detect and correct the error displacement in both $x$ and $p$ quadrature. Although it is prohibited by the uncertainty principle for both quadratures to be sharply defined, locally we can have a Wigner function with sharp peaks.  If we consider the Wigner function in Fig.\ \ref{fig:quadrature} and consider the center positive peak where all nearby positive regions are considered as a part of the peak, we have the variances in each quadrature as $\langle\Delta^2\hat{x}\rangle=1.45\pm0.03$ and $\langle\Delta^2\hat{p}\rangle=0.070\pm0.001$ and $\langle\Delta^2\hat{x}\rangle\langle\Delta^2\hat{p}\rangle=0.100\pm0.002$ which is well in the sub-Planck regime \cite{Zurek2001} ($\langle\Delta^2\hat{x}\rangle\langle\Delta^2\hat{p}\rangle<0.25$). The sub-Planck regime simultaneously in both quadratures is a regime that cannot be reached with simple Gaussian state, and this sub-Planck structure is a qualitative evidence that our approach can synthesize the sharp delta-function peak structure of the GKP state. Note that the value $\langle\Delta^2\hat{x}\rangle$ is limited by the amplitude of the initial kitten state and can be improved by using a cat state with a bigger amplitude.

Regarding the quantitative evaluation, we evaluate the stabilizers of the generated state. As we have mentioned, for our specific state $\ket{1_L}$, the two stabilizers become $\hat{S}_x$ and $\hat{S}_{\ket{1}}=-\hat{\bar{Z}}$. We calculate the average value of the stabilizers from the generated states and they are $\mean{\hat{S}_x}=0.170\pm0.003$ and $\mean{\hat{S}_{\ket{1}}}=0.216\pm0.006$, respectively. Although these values are still far from unity, they clearly surpass the values that can be achieved with classical states. Furthermore, their collective properties overcome the limits set by Gaussian states. See supplementary material for a more detailed analysis.

Although our work marks the first essential step toward generation of optical GKP states, there are still future technological improvements to be made. Our current limiting factor is the optical loss in the system. This optical loss can be lower in principle and the quality of the cat state can be much improved via techniques for generation of highly-pure cat states \cite{PhysRevA.96.052304,Asavanant:17}. On the other hand, we also need to make a cat state with larger amplitude and concatenate the process with deeper circuit depth as depicted in Fig.\ \ref{fig:schematic} to make high quality GKP states. Among many researches for generation of large-amplitude cat states, one of the most promising approaches would be the generalized photon subtraction \cite{PhysRevA.97.022341}, where the generated cat states are also squeezed. This removes the necessity of additional squeezing operation and makes the state more tolerant to losses \cite{PhysRevLett.120.073603}. Generalized photon subtraction also allows higher success rate than the conventional photon subtraction. Regarding the circuit depth, it is possible to use the time-domain multiplexing technique to realize a deep circuit depth in a hardware-efficient way \cite{PhysRevLett.118.110503}. Our current setup has the success rate of about 10 Hz which is not sufficient for actual computation. In addition to the aforementioned generalized photon subtraction, improving the squeezed light source and the photon number resolving detector can increase the success rate. Regarding the former, the improvement can be achieved by replacing the OPO in this work with our recent terahertz-bandwidth optical parametric oscillator \cite{10.1063/5.0063118}. Regarding the latter, although the photon counter in this experiment is based on SNSPD without photon number resolution, we have also shown that, with proper technique, SNSPD can be used for counting photons with high speed (timing jitter of  $\sim20$ ps) \cite{Endo:21}. Using these SNSPD would allow even higher generation rate of the cat state for GKP state generation. 

Finally, we briefly comment on how optical GKP states generated in the manner of this work can be used in actual quantum computation. As the GKP states generated in this method are a propagating wave, they can be coupled into the various types of optical quantum processors seamlessly without requiring any additional process such as wavelength conversion. As multimode linear operations \cite{PhysRevApplied.16.034005,Larsen2021}, quantum teleportation \cite{Furusawa706,Lee330}, and squeezing operation \cite{PhysRevLett.113.013601,PhysRevA.98.052311} on propagating optical field have already been demonstrated and implemented, all necessary operations on a GKP state, including syndrome measurement and recovery operation, can be readily applied. In particular, as the squeezed light sources and the homodyne detectors \cite{10.1063/5.0063118,10.1063/5.0137641} for the 5G telecommunication have been shown to be readily compatible with optical quantum computation, once the GKP states with sufficient quality are generated they can be immediately utilized in the ultra-high-speed quantum computation. Therefore, generation of optical GKP states in this paper demonstrates a viability and practicality of the propagating optical system as a venue for quantum computation.

\section*{Method}
The laser of our system is a continuous-wave laser with a wave length of 1545.32 nm and a second harmonic generator which generates 772.66 nm light. Our OPOs are a semi-monolithic cavity using a periodically poled KTiOPO$_{4}$ (PPKTP) crystal with a length of 10 mm. One side of the crystal acts as an output coupler and is coated with 90\% reflectivity at 1545 nm. The free spectral range of the cavity is 7.4 GHz. The homodyne detectors have a frequency bandwidth of about 200 MHz and the local oscillator power is set to about 3 mW. The interference visibility of the optical systems is on average above 96\%. Phase reference beams are injected for the phase stabilization. Each phase reference beam is detuned from the base frequency and the error signals for feedback controls are realized via beat notes similar to Ref.\ \cite{Asavanant:17,Asavanant373}. The SNSPD used in the photon subtraction for the generation of cat state is made of NbTiN and has an efficiency of about 75\% with dark counts of about 20 to 40 Hz. More details can be found in Ref.\ \cite{Miki:17}. To prevent the phase reference beam from reaching the SNSPDs, the sample\&hold method is employed where phase reference beams are blocked during the measurement (i.e., hold) phase and the lock point is hold during that period. The frequency filtering is done by using two bandpass filters put before the fiber coupling that leads to the SNSPD.

The count rate of the photon subtraction is about 80-90 kHz for both OPOs with the fake count of about 0.5-1.0 kHz. As no quantum memories are employed in this experiment, the interference of the cat states is considered successful when the time difference between two SNSPDs is within $\pm0.6$ ns. The mode match between two wave packets when there is 0.6 ns time difference is above 93\%. The success rate of the interference is about 10 Hz and the homodyne conditioning is done for the quadrature in the range of $\pm0.3$, which corresponds to about 30\% for the current experimental parameters.

Quantum state generation is implemented using the maximum likelihood method \cite{Lvovsky_2004}. The measurements are done for tthe six phases of the quadrature ($0^\circ$, $30^\circ$, $60^\circ$, $90^\circ$, $120^\circ$, $150^\circ$) and the number of the measurements are determined so that after the conditioning, the number of data points for each phase is about 20,000. 

\bibliography{ref.bib}

\section*{Acknowledgments}
This work was partly supported by Japan Science and Technology (JST) Agency (Moonshot R \& D) Grant No. JPMJMS2064 and JPMJMS2066, UTokyo Foundation, and donations from Nichia Corporation of Japan. W.A. acknowledge the funding from Japan Society for the Promotion of Science KAKENHI (No. 23K13040). M. E. acknowledge the funding from JST (JPMJPR2254). W.A. and M.E. acknowledge supports from Research Foundation for OptoScience and Technology. P.v.L. acknowledges funding from the BMBF in Germany (QR.X, PhotonQ, QuKuK), from the EU/BMBF via QuantERA (ShoQC) and from the EU's HORIZON Research and Innovation Actions (CLUSTEC). P.M. acknowledges Grant No. 22-08772S of Czech Science Foundation (GACR) and the European Union’s HORIZON Research and Innovation Actions under Grant Agreement no. 101080173 (CLUSTEC). R.F. acknowledges the project 21-13265X of Czech Science Foundation. P.M. and R.F. acknowledges EU H2020-WIDESPREAD-2020-5 project NONGAUSS (951737) under the CSA - Coordination and Support Action.

\section*{Author contributions}
S.K. leads the experiment with the supervision from W.A., K.T., M.E. and A.F.. S.K. collects the experimental data. S.K. and W.A. analyze the data. W.A. visualizes the data for the manuscript. Theoretical discussions and interpretations of the data is done by W.A.,  P.v.L., R.F., P.M., F.H., H.N., K.F., A.S., R.I.. The criteria in the supplementary is developed by P.M., W.A., and R.F.. H.T., S.M., M.Y., and F.C. provide the SNSPD used in this experiment. W.A. writes the manuscript and the supplementary material with the helps of P.v.L., A.F., P.M., R.F., and the other authors.

\section*{Additional information}
For correspondence, contact A.F. or W.A..

\section*{Competing financial interests}
The authors declare no competing financial interests.


\widetext
\begin{center}
\textbf{\large Supplementary material for ``Propagating Gottesman-Kitaev-Preskill states encoded in an optical oscillator''}
\end{center}

\section{Experimental setup}
\begin{figure*}[htp]
\includegraphics[width=\textwidth]{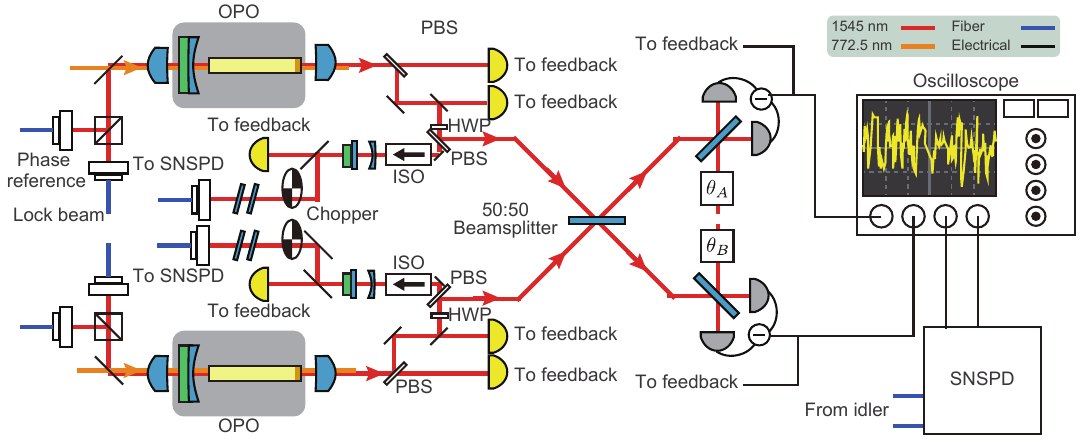}%
\caption{The detailed optical setup. Note that the reference cavity for the pump beam, the setup for beams detuning, and the circuitry for feedback controls are omitted from this figure. OPO, optical parametric oscillator; PBS, polarization beamsplitter; SNSPD, superconducting nanostripped single photon detector; HWP, half-wave plate; ISO, isolator.\label{fig_S:schematic}}
\end{figure*}
Figure \ref{fig_S:schematic} shows the detailed experimental setup. In this setup 1545.32 nm continuous-wave laser (NKT Photonics) is used as a master laser. The 772.66-nm pump beam is generated using an second-harmonic-generator module. Around 2 W of 1545.32-nm light and 500 mW of 772.66-nm light are used in the whole experiment. The 1545.32-nm light is split and used for four main purposes: the local oscillators (LOs) of the homodyne detectors, the alignment beams of the optical parametric oscillators (OPOs), the phase reference beams of the OPOs, and the lock beams of the optical cavities. For the stabilization, the phase reference beam and the lock beam are detuned from the fundamental frequency. The lock beam is detuned by 2 MHz, while the phase reference beam for each OPO is detuned by 1.14 MHz and 1.43 MHz, respectively.

The two OPOs are semi-monolithic cavity with a 10-mm periodically-poled KTP (PPKTP) (Raicol) crystal placed inside. The concave mirror has a radius of curvature of 8 mm and is coated with high-reflectivity coating at 1545.32 nm and anti-reflection coating at 772.66 nm. The backside of the mirror also have AR coat at 1545.32 nm and 772.66 nm. The PPKTP crystal is a type-0 crystal and have an AR coat at 1545.32 nm and 772.66 nm on the surface near the concave mirror. On the other surface 90\% reflectivity coating at 1545.32 nm and AR coating at 772.66 nm is implemented. The mechanical structure is based on Ref.\ \cite{Takanashi:19}. To stabilize the cavity Pound-Drever-Hall locking using the lock beam is used and the modulation frequency is 29MHz. For the pump beam, as it does not resonate with the OPO, an additional optical cavity (omitted from Fig.\ \ref{fig_S:schematic}) is used to match the spatial mode of the pump beam and of the OPO.

The idler path for the detection of the subtracted photon consists of a Fabry-Perot cavity and a band-pass filter. The idler path is fiber coupled and the photon is detected by the superconducting nanostrip photon detector (SNSPD) \cite{Miki:17}. Mechanical shutter is used for blocking the lights from entering the SNSPD. The throughput efficiency of the idler path until the fiber coupler is about 75\% for both OPO. The transmission, including the coupling efficiency, of the fiber leading to the SNSPD is about 95\%. The efficiency and the dark count of the SNSPD is about 75\% and 25-40 Hz for both SNSPDs, respectively. This results in about the total of 50\% efficiency on the idler path.

The visibility for all the beam splitter in the system is above 96\%. The phase reference beams are injected into each OPO for the stabilization of the relative phase at each beamsplitter and the relative phase between the pump beams and the LOs. The error signals for the feedback control are obtained by detecting the beat note from the interference of the detuned phase reference beams. This experiment utilizes sample \& hold method where the lights for the stabilization and control of the systems are chopped and the feedback controls are held during the data aquisitions. The phase reference lights are switched on-off using the AOMs and there is a mechanical chopper in front of the fiber coupler leading to the SNSPDs to prevent strong phase reference lights from reaching the SNSPDs.

The homodyne detectors used in this experiment has a bandwidth of about 200 MHz and the LO powers are set to about 3 mW with the stabilization system (Thorlabs). The signals of the homodyne detectors are recorded using an oscilloscope whose trigger is the the signal of the two SNSPDs, where we limit the time difference between the arrival of each SNSPD signal to be $\pm0.6$ ns. This value is determined from the shape of the temporal mode of the generated cat state in each path. The maximum mode matching is 94.6\% and even with $\pm0.6$ ns shift, the mode matching stays higher than 93\%.

\section{Data analysis}
Of the two homodyne detectors, HD1 is used for the conditioning, so the measurement basis is fixed to $x$. On the other hand, the phase of HD2 is set to $0^\circ$, $30^\circ$, $60^\circ$, $90^\circ$, $120^\circ$, and $150^\circ$ for collecting the quadrature data to do the state reconstruction with quantum tomography. The number of the events is selected so that after conditioning on the measurement results of HD1, the number of the conditioned events is about 10000 events for each phase. The window for the conditioning at HD1 is for the quadrature values of $\pm0.3$, resulting in about 30\% success probability in the current setup.

We reconstruct the density matrix in the Fock basis using the acquired quadrature values via the maximum likelihood method \cite{Lvovsky_2004}. To correct for Gaussian operation, we calculate the required Gaussian correction by maximizing the fidelity of the Gaussian corrected state to the nearest approximated GKP state. The maximum fidelity is $0.551\pm0.003$ to the approximated GKP state of squeezing level of about 2.5 dB. Using the Gaussian corrected density matrix, Wigner function can be calculated and the values of the various stabilizers are calculated by the Fourier transform of the Wigner function. The error bars of the reported results in this paper are calculated using bootstrapping method.

\section{Gaussian limit criteria for GKP stabilizer}
Here we discuss the classical limit criteria for the GKP stabilizer. For simplicity, let us define the displacement operator in phase space $\hat{D}(x_0,p_0)$ as the displacement by $(x_0,p_0)$ in the phase space of the Wigner function. They can be related to the stabilizer in the main text as $\hat{S}_x = \hat{D}(2\sqrt{\pi},0)$, $\hat{S}_p = \hat{D}(0,2\sqrt{\pi})$, and $\hat{S}_{\ket{1}}=-\hat{D}(0,\sqrt{\pi})$. Note that as displacement in $x$ and $p$ does not commute, there could be an extra phase factor, but as they do not change the norm, we will ignore it here without loss of generality.

Here we will consider two types of criteria given by
\begin{align}
C_1(g)&=\abs{\mean{\hat{D}(2g\sqrt{\pi},0)}\mean{\hat{D}(0,\sqrt{\pi}/g)}}\\
C_2(g)&=\abs{\mean{\hat{D}(2g\sqrt{\pi},0)+\hat{D}(-2g\sqrt{\pi},0)+\hat{D}(0,2\sqrt{\pi}/g)+\hat{D}(0,-2\sqrt{\pi}/g)}}
\end{align}
where $0<g$ is the parameter of effective squeezing that we can apply to the quantum state in order to better fit the position of the peaks to the GKP grid. First, we can show that for any coherent state $\ket{\alpha}$, 
\begin{align}
\bra{\alpha}\hat{D}(x_0,p_0)\ket{\alpha}=\bra{0}\hat{D}(x_0,p_0)\ket{0}
\end{align}
This means that for any classical mixture represented by a coherent state, the values of the above criteria (which we will denote as $C_{1,\textrm{cl}}(g)$ and $C_{2,\textrm{cl}}(g)$) is equal to that of a coherent state. Then, we can calculate $C_{1,\textrm{cl}}(g)$ as 
\begin{align}
C_{1,\textrm{cl}}(g)=\abs{\bra{0}\hat{D}(2g\sqrt{\pi},0)\ket{0}\bra{0}\hat{D}(0,\sqrt{\pi}/g)\ket{0}}
\end{align}
For coherent state, at $\hbar=1$ we have
\begin{align}
\alpha&=\frac{1}{\sqrt{2}}(x_0+ip_0)\\
\langle\alpha\vert 0 \rangle&=\exp\left(-\frac{\abs{\alpha}^2}{2}\right)
\end{align}
Combining everything, we arrive at
\begin{align}
\begin{split}
C_{1,\textrm{cl}}(g)&=\exp\left(-\frac{\abs{g\sqrt{2\pi}}^2+\abs{i\sqrt{\frac{\pi}{2}}/g}^2}{2}\right)\\
&=\exp\left[-\frac{\pi}{2}\left(2g^2+\frac{1}{2g^2}\right)\right]
\end{split}
\end{align}
For the classical state, the maximum of $C_{1,\textrm{cl}}(g)$ (which we will denote as $\mathcal{C}_{1,\textrm{cl}}$) is achieved when $2g^2=1$ which corresponds to
\begin{align}
\mathcal{C}_{1,\textrm{cl}}=\exp(-\pi)
\end{align}
similar classical threshold can also be found for $C_2(g)$ which gives
\begin{align}
C_{2,\textrm{cl}}(g)&=2\left[\exp\left(-g^2\pi\right)+\exp\left(-\frac{\pi}{g^2}\right)\right]\\
\mathcal{C}_{2,\textrm{cl}}&=2
\end{align}
The maximum of $C_{2,\textrm{cl}}(g)$ is attained at $g\to0$ or $g\to\infty$. This corresponds to the infinitely squeezing or antisqueezing the state. These maximum values indicate that if we perform experimental state generation and measure the stabilizer to be more than these values, at the very least, the generated state surpass the classical limitation. 

Next we consider whether these criteria also holds for the arbitrary Gaussian states including mixed states. First we notice that squeezing in $x$ or $p$ quadrature corresponds to simply changing the value of $g$, meaning that the maximum does not change even with the squeezing in these two directions. Next let us consider rotation given by operator $\hat{R}(\theta)$. Arbitrary Gaussian pure states can be written as
\begin{align}
\ket{\psi(\alpha,\theta,r)}=\hat{D}(\alpha)\hat{R}(\theta)\hat{S}_{q}(r)\ket{0},
\end{align}
with $\hat{S}_{q}(r)$ being a squeezing operator transforming the quadrature as $\hat{x}\to\exp(-r)\hat{x}$ and $\hat{p}\to\exp(r)\hat{p}$. The rotation transforms the displacement operator as
\begin{align}
\hat{R}^\dagger(\theta)\hat{D}(x_0,p_0)\hat{R}(\theta)=\hat{D}(\cos\theta x_0+\sin\theta p_0,\cos\theta p_0-\sin\theta x_0)
\end{align}
Therefore, $C_{1}(g)$ for Gaussian states becomes
\begin{align}
\begin{split}
C_{1,\textrm{Gaussian}}(g,\theta,r)&=\abs{\bra{0}\hat{S}_{q}^\dagger(r)\hat{D}(2g\sqrt{\pi}\cos\theta,-2g\sqrt{\pi}\sin\theta)\hat{S}_{q}(r)\ket{0}\bra{0}\hat{S}_{q}^\dagger(r)\hat{D}(\sqrt{\pi}\sin\theta/g,\sqrt{\pi}\cos\theta/g)\hat{S}_{q}(r)\ket{0}}\\
&=\abs{\bra{0}\hat{D}(2g\exp(r)\sqrt{\pi}\cos\theta,-2g\exp(-r)\sqrt{\pi}\sin\theta)\ket{0}\bra{0}\hat{D}(\sqrt{\pi}\exp(r)\sin\theta/g,\exp(-r)\sqrt{\pi}\cos\theta/g)\ket{0}}\\
&=\exp\left[-\frac{\pi}{2}\left([\exp(2r)\cos^2\theta+\exp(-2r)\sin^2\theta]2g^2+[\exp(2r)\sin^2\theta+\exp(-2r)\cos^2\theta]\frac{1}{2g^2}\right)\right]
\end{split}
\end{align}
From the above equation, we can find the maximum with respect to $g$ as
\begin{align}
\begin{split}
\mathcal{C}_{1}(\theta,r)&=\exp\left[-\pi\sqrt{\cos^4\theta+\sin^4\theta+2\cosh4r\sin^2\theta\cos^2\theta}\right]\\t
&=\exp\left[-\pi\sqrt{1+\sinh^2 2r\sin^22\theta}\right]
\end{split}
\end{align}
We can see that, regardless of the values of $\theta$, the maximum occurs when $r=0$ and coincides with $\mathcal{C}_{1,\textrm{cl}}$ meaning that this is also the Gaussian limit.

Similar calculation for $C_{2}$ gives
\begin{align}
C_{2,\textrm{Gaussian}}^{\textrm{bound}}(g,\theta,r)&=2\left[\exp\left(-\exp(2r)\cos^2\theta+\exp(-2r)\sin^2\theta]g^2\pi\right)+\exp\left(-[\exp(2r)\sin^2\theta+\exp(-2r)\cos^2\theta]\frac{\pi}{g^2}\right)\right]\\
\mathcal{C}_{2}(\theta,r)&=2
\end{align}
which coincide with the classical limit. Note that unlike the classical criteria, as we have varied $g$ first,  the Gaussian limits, in principle, does not depend on $g$ and thus act as a stronger criteria than the classical limit.

\begin{figure*}[htp]
\includegraphics[width=0.7\textwidth]{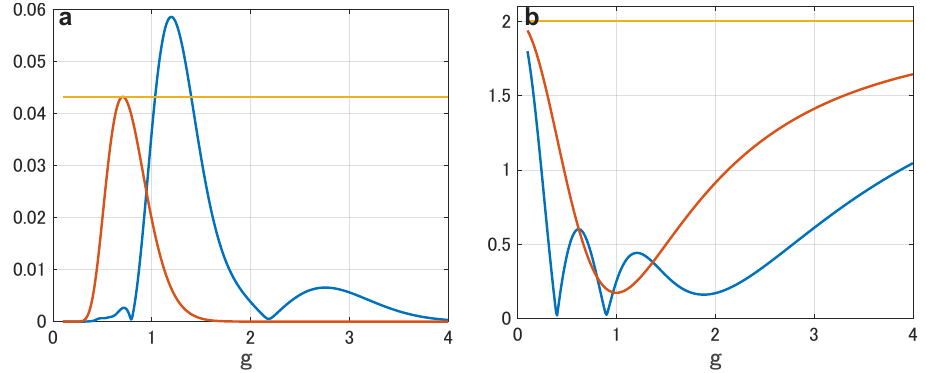}%
\caption{Plot of the criteria. (A) $C_1$. (B) $C_2$. The blue lines are the calculated $C_1(g)$ ($C_2(g)$) from the experimental results. The orange lines are plot of the classical limit given by $C_{1,\textrm{cl}}(g)$ ($C_{2,\textrm{cl}}(g)$). The yellow line is the plot of the Gaussian limit given by the maximum of $\mathcal{C}_1(\theta,r)$ ($\mathcal{C}_2(\theta,r)$)with respecct to $\theta$ and $r$. Note that the Wigner function used here is the Wigner function corrected for the Gaussian operations.\label{fig_S:criteria}}
\end{figure*}

When applying the criteria for the experimental results (which we will put as $C_{\textrm{exp}}$), if there is a value $g$ such that $C_\textrm{exp}(g)/C_{\textrm{cl}}(g)>1$, the state has nonclassical GKP properties. On the other hand, if there is a value of $g$ such that $C_\textrm{exp}(g)/\textrm{max}_{g}[C_\textrm{cl}(g)]>1$, the state exhibits non-Gaussian GKP properties. The results for applying these criteria are shown in Fig.\ \ref{fig_S:criteria}. We observe that for $C_1$, both classical and Gaussian limits are violated, while for $C_2$, only classical limit is violated.

\end{document}